\documentclass[twocolumn]{aastex631}
\usepackage{multirow}
\usepackage[tbtags]{amsmath}

\newcommand{\gr}{$\gamma$-ray}

\graphicspath{{./}{figures/}}

\begin{document}

\title{Discovery of $\gamma$-Ray Pulsations from the Extreme-Spin-Down Millisecond Pulsar PSR J0435+3233}

\author{Mengqing Zhang}
\affiliation{Department of Astronomy, School of Physics and Astronomy, Key Laboratory of Astroparticle Physics of Yunnan Province, Yunnan University, Kunming 650091, People's Republic of China; zhangpengfei@ynu.edu.cn}

\author{Shengbin Pei}
\affiliation{Department of Astronomy, School of Physics and Astronomy, Key Laboratory of Astroparticle Physics of Yunnan Province, Yunnan University, Kunming 650091, People's Republic of China; zhangpengfei@ynu.edu.cn}

\author{Pengfei Zhang}
\affiliation{Department of Astronomy, School of Physics and Astronomy, Key Laboratory of Astroparticle Physics of Yunnan Province, Yunnan University, Kunming 650091, People's Republic of China; zhangpengfei@ynu.edu.cn}

\begin{abstract}

Motivated by the recent discovery of PSR~J0435+3233, a millisecond pulsar with an
exceptionally large period derivative of $\dot{P}=4.9\times10^{-17}\ {\rm s\,s^{-1}}$
and a high spin-down luminosity of $\dot{E}=5.89\times10^{37}\ {\rm erg\,s^{-1}}$,
we analyze $\sim$17.7~yr of observations obtained with the \textit{Fermi} Large Area Telescope for the pulsar.
We identify the cataloged source 4FGL~J0435.5+3232, located only $0.01^{\circ}$
from the radio timing position, as the $\gamma$-ray counterpart of PSR~J0435+3233.
Using $\gamma$-ray events collected within the validity interval of the radio timing
ephemeris, we detect its $\gamma$-ray pulsations at a $\sim6.8\sigma$ confidence level.
In this interval, the phase-resolved analysis shows that the pulsed emission is
concentrated predominantly within the rotational-phase of $\phi\sim0.44$--$0.69$.
We also derive its $\gamma$-ray luminosity of $L_{\gamma}=6.26\times10^{32}\ {\rm erg\,s^{-1}}$,
assuming a distance of 1.2~kpc and isotropic emission. This luminosity corresponds
to an apparent $\gamma$-ray efficiency of $\eta_{\gamma}\sim1.1\times10^{-5}$,
revealing an exceptionally low $\gamma$-ray output despite the pulsar's young-pulsar-like
rotational-energy budget. Our detection establishes PSR~J0435+3233 as
a $\gamma$-ray MSP, and the striking combination of its high spin-down power and
low apparent $\gamma$-ray efficiency provides a new probe of particle acceleration,
radiation beaming, and viewing geometry in the magnetospheres of millisecond pulsars
with extreme rotational properties.

\end{abstract}

\keywords{Gamma-ray sources(633); Gamma-rays(637); Millisecond pulsars (1062); Pulsars (1306)}

\section{Introduction}
\label{Intro}


Millisecond pulsars (MSPs) are rapidly rotating neutron stars with spin periods of
typically a few milliseconds \citep{bkh+82}. In the $P$--$\dot{P}$ diagram, they
form a population distinct from normal pulsars, characterized by short spin periods
and, in most cases, small period derivatives \citep{bv91}.
Approximately three quarters of known MSPs reside in binary systems\citep{Manchester2005},
and most are believed to have been spun up through a
recycling process, during which the accretion of matter and angular momentum from
a companion star accelerates the neutron star to millisecond
periods \citep{bkh+82,Manchester2005,bv91,l08}. This evolutionary scenario is strongly
supported by accreting millisecond X-ray pulsars, which provide direct evidence for
accretion-driven spin-up \citep{wv98,pw21}, and by transitional MSPs that alternate between
rotation-powered radio-pulsar states and accretion-powered X-ray
states \citep{asr+09,pfb+13,jah+16}. Together, these systems provide an observational link
between accreting neutron stars in X-ray binaries and recycled radio MSPs.

Rotation-powered MSPs can convert a fraction of their spin-down power into high-energy
radiation through particle acceleration in their magnetospheres.
Observations with the \textit{Fermi} Large Area Telescope (\textit{Fermi}-LAT) have
established MSPs as an important population of Galactic GeV sources, many of which
exhibit pulsations at their spin periods and curved spectra with cutoffs at
GeV energies \citep{2pc,3pc}. The detection of $\gamma$-ray pulsations provides
an unambiguous association between a $\gamma$-ray source and a pulsar,
while the pulse profile, spectrum, and luminosity offer important diagnostics of
magnetospheric particle acceleration, radiation geometry, and energy-conversion
efficiency. Pulsars with large spin-down luminosities and high spin-down fluxes,
$\dot{E}/d^{2}$, are therefore particularly promising targets for searches for
pulsed GeV emission \citep{3pc}.

Recently, \citet{wwy+26na} reported the discovery of PSR~J0435+3233 with FAST.
This binary MSP has a spin period of $P=3.20$~ms and an exceptionally large
period derivative of $\dot{P}=4.88\times10^{-17}\ {\rm s\,s^{-1}}$, at least two
orders of magnitude larger than those of other known MSPs. It consequently
occupies a distinctive region of the $P$--$\dot{P}$ diagram, well above
the classical spin-up line. Its unusual rotational properties are difficult to
reconcile with the standard binary-recycling scenario and pose a significant
challenge to current theories of MSP formation and accretion-driven spin-up.
Possible explanations include super-Eddington accretion onto a strongly magnetized
neutron star and the accretion-induced collapse of
a magnetized oxygen--neon--magnesium white dwarf \citep{wwy+26na}. PSR~J0435+3233
also has an exceptionally high spin-down luminosity
of $\dot{E}=5.89\times10^{37}\ {\rm erg\,s^{-1}}$, comparable to those of young,
energetic pulsars. Together with its relatively small DM-inferred distance of
$\sim$1.2~kpc, this large rotational energy budget makes PSR~J0435+3233
a particularly promising candidate for detectable $\gamma$-ray emission.

Motivated by these properties, we analyze $\sim$17.7~yr of \textit{Fermi}-LAT observations in
the 0.1--500~GeV energy range for this pulsar. The cataloged $\gamma$-ray source 4FGL~J0435.5+3232,
listed in the Fourth \textit{Fermi}-LAT Source Catalog Data Release~4
\citep[4FGL-DR4;][]{4fgl-dr4}, lies only $0.01^{\circ}$ from the radio timing position
of PSR~J0435+3233. Using LAT data collected within the ephemeris-valid interval (EVI),
we detect pulsed $\gamma$-ray emission from the MSP at a $\sim6.8\sigma$ confidence level.
The positional coincidence, pulsations at the spin period, and phase-dependent $\gamma$-ray
emission collectively establish 4FGL~J0435.5+3232 as the $\gamma$-ray counterpart of
PSR~J0435+3233.
Here, we present our LAT analysis and discuss the high-energy emission properties
and physical implications of this extreme MSP.

\section{Data Analysis and Results}
\label{sec:lat-data}

\subsection{Data Reduction and Source Model}
\label{sec:data}

We analyzed Pass~8 \textit{Fermi}-LAT SOURCE-class events (\texttt{evclass = 128}),
including both FRONT- and BACK-converting events (\texttt{evtype = 3}),
in the 0.1--500.0~GeV energy range. The data span the period from 2008 August 4 to
2026 May 1. Events were selected from a $20^{\circ}\times20^{\circ}$ region of
interest (ROI) centered on the radio timing position of PSR~J0435+3233
(R.A.=$04^{\mathrm{h}}35^{\mathrm{m}}33^{\mathrm{s}}.76$, decl.=$+32^{\circ}33^{\prime}07^{\prime\prime}.93$).
To reduce contamination from
$\gamma$-rays produced in the Earth's limb, we excluded events with zenith angles
greater than $90^{\circ}$. We further retained only data collected during standard
science operations by applying the quality filter \texttt{DATA\_QUAL>0 \&\& LAT\_CONFIG==1}.
The analysis was performed using Fermitools version~2.2.0 and the \texttt{P8R3\_SOURCE\_V3}
instrument response functions.

Using the $\sim$17.7~yr \textit{Fermi}-LAT data, we constructed the initial source model
based on 4FGL-DR4 using the \texttt{make4FGLxml.py} script and optimized the model parameters
through a binned maximum-likelihood analysis. The $\gamma$-ray emission from
4FGL~J0435.5$+$3232 was initially modeled with a log-parabola (LP) spectrum,
$\frac{\mathrm{d}N}{\mathrm{d}E}=N_0(\frac{E}{E_b})^{-[\alpha+\beta\ln(E/E_b)]}$.
During the likelihood fit, the normalization and spectral-shape parameters of all cataloged
$\gamma$-ray sources within $5^{\circ}$ of 4FGL~J0435.5$+$3232 were allowed to vary.
For sources located between $5^{\circ}$ and $10^{\circ}$ from the target, only their
normalizations were left free. The normalizations of sources beyond $10^{\circ}$ that are
flagged as variable in 4FGL-DR4, as well as those of the Galactic and
isotropic diffuse-emission components, were also allowed to vary. All remaining model
parameters were fixed at their 4FGL-DR4 values.

Given the association of the $\gamma$-ray source with PSR~J0435+3233, we subsequently
refitted 4FGL~J0435.5$+$3232 using a power law with a new subexponentially cutoff power law (PLEC4),
a spectral form commonly used to describe magnetospheric $\gamma$-ray emission from
pulsars,
$\frac{\mathrm{d}N}{\mathrm{d}E}=A(\frac{E}{E_0})^{\Gamma_0+d/b}e^{\frac{d}{b^2}(1-\left(\frac{E}{E_0}\right)^b)}$.
The best-fit spectral parameters are summarized in Table~\ref{tab:par}.
Given its physical applicability to pulsar magnetospheric emission and its slightly
higher TS value, we adopted the PLEC4 model for all subsequent analyses.
Unless otherwise stated, all quoted uncertainties, including those reported hereafter,
correspond to $1\sigma$ statistical errors.


\subsection{Timing Analysis}
\label{sec:tim}

For the pulsation analysis, we selected photons in the 0.1--500~GeV within $3.16^{\circ}$
of 4FGL~J0435.5$+$3232 \citep{aaa+10}. Using the best-fit PLEC4 source model described above,
we calculated the probability that each photon originated from the target source with
\texttt{gtsrcprob} and adopted the resulting source probabilities as photon weights.
Photon arrival times were assigned rotational phases using \textsc{Tempo2} \citep{hem06}
and its \textit{Fermi} plug-in \citep{rkp+11}, together with the radio timing solution
reported by \citet{wwy+26na}. The significance of the pulsed signal was evaluated using
the weighted H-test \citep{jb10,k11}. We first performed an exploratory folding analysis
using the whole LAT data collected between 2008~August~4 and 2026~May~1, which yielded
a weighted H-test statistic of only H-value$\sim9.4$. The radio timing solution, however,
is valid only over MJD~59101--60898 \citep{wwy+26na}, and extrapolating it beyond
this interval cannot reliably preserve rotational phase coherence. The low H-test value
obtained from the whole data is therefore consistent with phase decoherence outside
the EVI. We consequently restricted the pulsation analysis to events collected within MJD~59101--60898.

\begin{figure*}
\centering
\includegraphics[angle=0,scale=0.55]{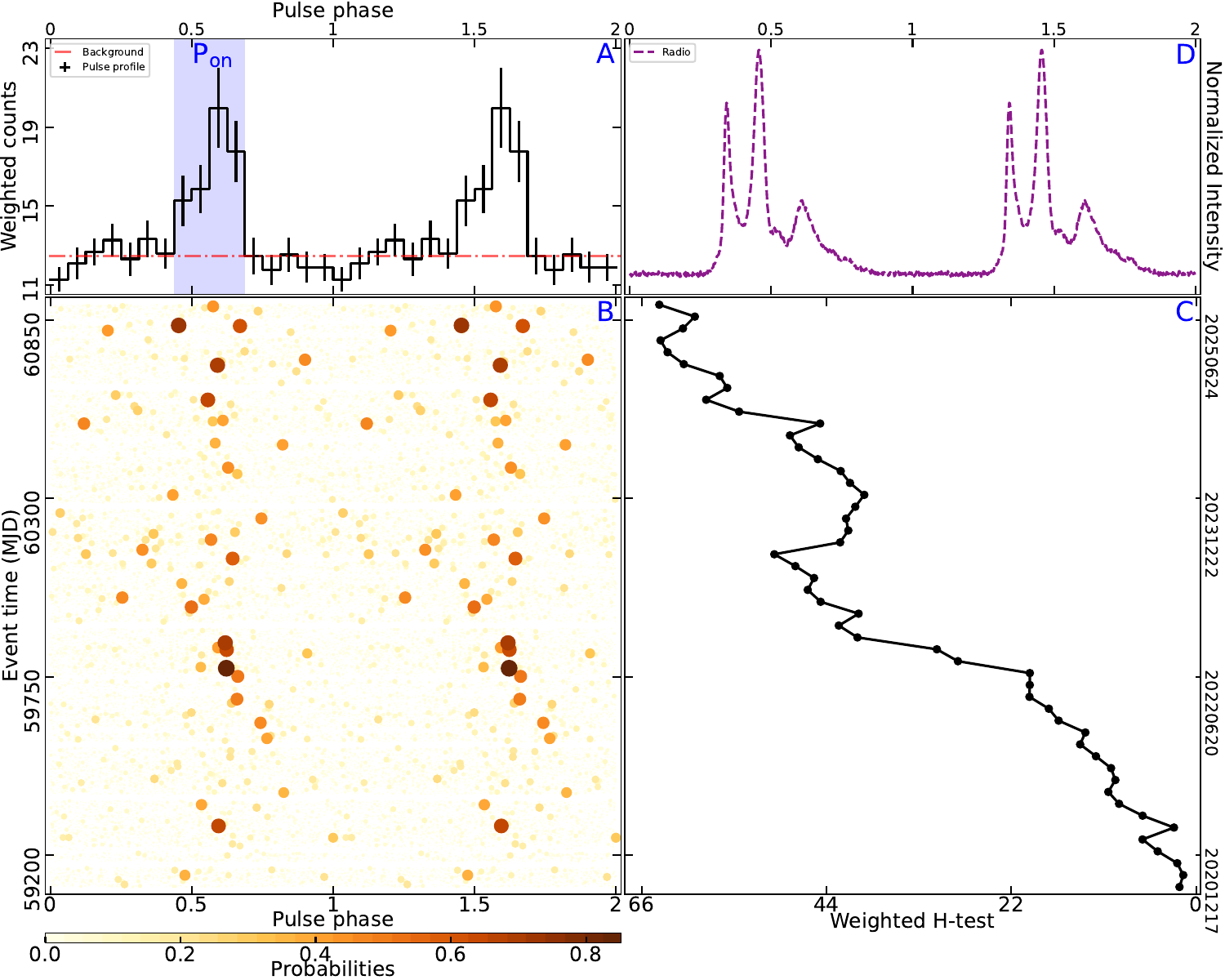}
\caption{\textit{Fermi}-LAT timing analysis of target over the EVI. (A) Integrated weighted
         \gr\ pulse profile. The horizontal dashed line indicates the estimated
         background level. (B) Two-dimensional phase--time diagram of
         the weighted photons. The color scale represents the probability weight
         assigned to each photon, with a maximum weight of $\sim$0.85.
         (C) Cumulative weighted H-test statistic as a function of time.
         (D) Schematic radio pulse profile reconstructed from that
         reported by \citet{wwy+26na} for qualitative comparison.
         Two rotational cycles are shown in panels~(A), (B), and (D) for clarity.}
\label{fig:timing}
\end{figure*}

Applying the same phase-assignment and weighting procedures to the EVI data,
we constructed the weighted pulse profile and the two-dimensional phase--time diagram
shown in Figures~\ref{fig:timing}(A) and (B), respectively.
The pulse profile was divided into 16 rotational-phase bins, and the uncertainty in
each bin was calculated following the weighted-profile prescription of \citet{2pc}.
The $\gamma$-ray profile exhibits a prominent peak within the phase interval
$\phi=0.4375$--$0.6875$. We therefore defined this interval as the on-pulse region and
the remaining phases as the off-pulse region for the subsequent phase-resolved analyses.
A schematic radio pulse profile reconstructed from that reported by \citet{wwy+26na}
is shown in Figure~\ref{fig:timing}(D) for qualitative comparison. The radio pulse
appears to lead the $\gamma$-ray pulse in the displayed profiles. However, because the
two profiles have not been absolutely phase-aligned, the apparent phase offset cannot
be reliably quantified, precluding a detailed physical interpretation of
the radio-to-$\gamma$-ray phase lag. Following \citet{2pc},
we estimated an average background level of 12.5 weighted counts per phase bin,
which is consistent with the counts in off-pulse.

Figure~\ref{fig:timing}(C) shows the cumulative weighted H-test statistic over the EVI.
The statistic generally increases as additional LAT data are accumulated, reaching
H-test value $\sim63.9$ for the full interval. According to the null distribution of the
weighted H-test, this value corresponds to a chance probability of $p\simeq7.6\times10^{-12}$,
or a detection significance of $\sim6.8\sigma$.

\begin{table}
\begin{center}
\caption{Results of Likelihood analysis}
\begin{tabular}{ccccc}
\hline\hline
Models & \multicolumn{4}{c}{Best-fit Parameters}  \\
\hline
LP     &  TS & $\alpha$ & $\beta$ & $E_b$~(GeV) \\
(Whole) &  99.4 & 2.30(15) & 0.18(4) & 2825.8 \\
\hline
PLEC4   &  TS & $\Gamma$ & $d$ & $b$ \\
(Whole) &  100.5 & 1.93(20) & 0.17(9) & 2/3 \\
(EVI)   &  21.3 & 2.12(33) & 0.13(9) & 2/3 \\
\hline
Phase   & \multicolumn{4}{c}{Phase-resolved analysis}  \\
\hline
On-pulse  & 77.0 & 1.93(18) & 0.14(10) & 2/3 \\
Off-pulse & 0.0  & 2.12$^\mathrm{a}$ & 0.13$^\mathrm{a}$ & 2/3 \\
\hline
\end{tabular}
\label{tab:par}
\end{center}
{\bf Notes. }{Best-fit results of target for the whole data, the EVI, and the on-pulse
              and off-pulse phase selections. Numbers in parentheses denote the $1\sigma$
              statistical uncertainties in the last quoted digits. Parameters reported
              without uncertainties were fixed at their corresponding 4FGL-DR4 values,
              except in the off-pulse analysis, for which they were fixed
              at the best-fit values obtained from the EVI data.}
\end{table}

\begin{figure}
\centering
\includegraphics[angle=0,scale=0.55]{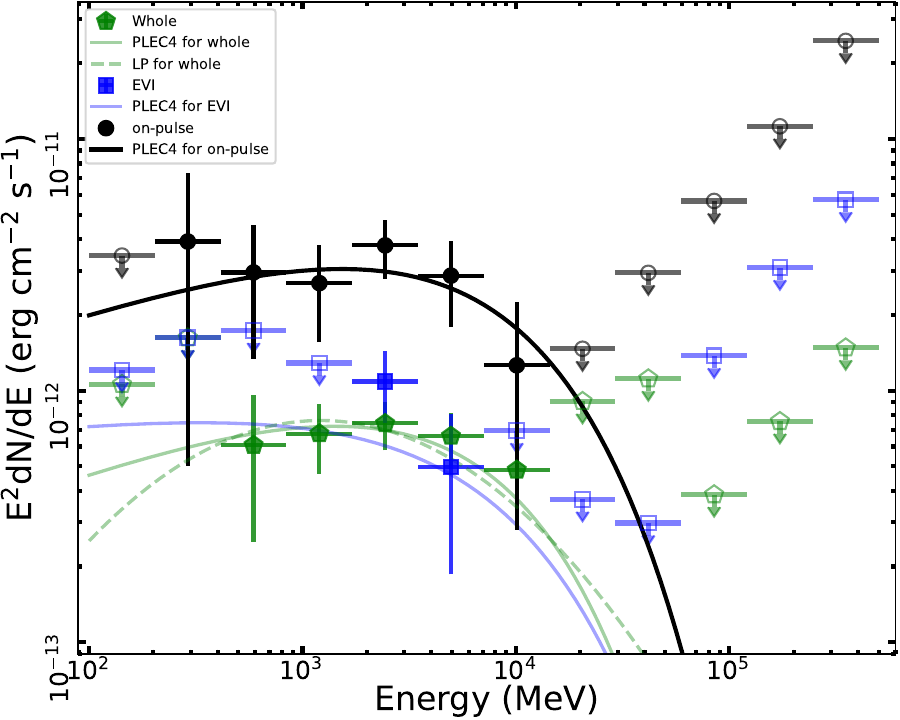}
\caption{$\gamma$-ray SEDs of target obtained from the whole data, the EVI, and
         the on-pulse selection within the EVI, shown in green, blue, and black, respectively.
         The best-fit PLEC4 and LP models are represented by the solid and dashed curves,
         respectively. Flux measurements are shown for energy bins with $\mathrm{TS}\geqslant4$,
         whereas 95\% upper limits are given for bins with $\mathrm{TS}<4$.}
\label{fig:sed}
\end{figure}

\subsection{Spectral and Phase-resolved Analysis}
\label{sec:phlc}

To characterize the $\gamma$-ray spectral energy distribution (SED) of the target, we first
performed a likelihood analysis using the whole \textit{Fermi}-LAT data.
To construct the SED, we divided the energy range 0.1--500.0~GeV into 12 equally logarithmically
spaced bins and performed an independent likelihood analysis in each bin using the best-fit model
described above, the target was modeled with the PLEC4 model,
which is shown as the green solid curve in Figure~\ref{fig:sed}. The resulting SED measurements
are shown as the green data points in Figure~\ref{fig:sed}.

For comparison, we repeated the phase-averaged spectral analysis using all events collected
within the EVI. We then performed phase-resolved spectral analyses using the data from
the on-pulse and off-pulse intervals defined in Section~\ref{sec:tim}. The best-fit spectral
parameters and them obtained from the whole data, the EVI data, and the on-pulse and off-pulse
selections are summarized in Table~\ref{tab:par}.
Their corresponding SEDs and best-fit spectral models are shown in Figure~\ref{fig:sed}.
For the phase-averaged results within the EVI, we obtained an integrated 0.1--500.0~GeV
energy flux of $G_{\gamma}=(3.63\pm1.44)\times10^{-12}\  {\rm erg\,cm^{-2}\,s^{-1}}$.

\begin{figure*}
\centering
\includegraphics[angle=0,scale=0.6]{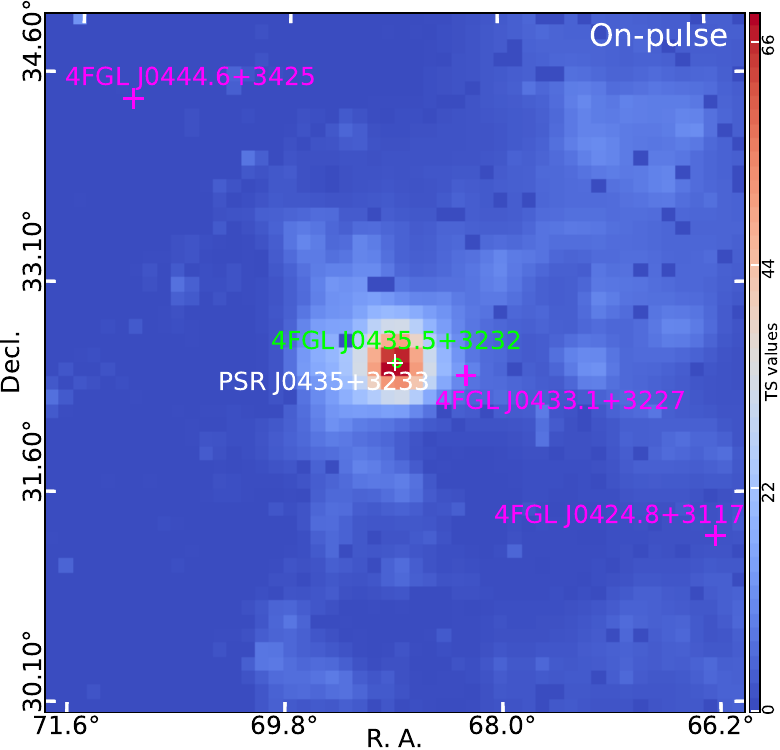}
\includegraphics[angle=0,scale=0.6]{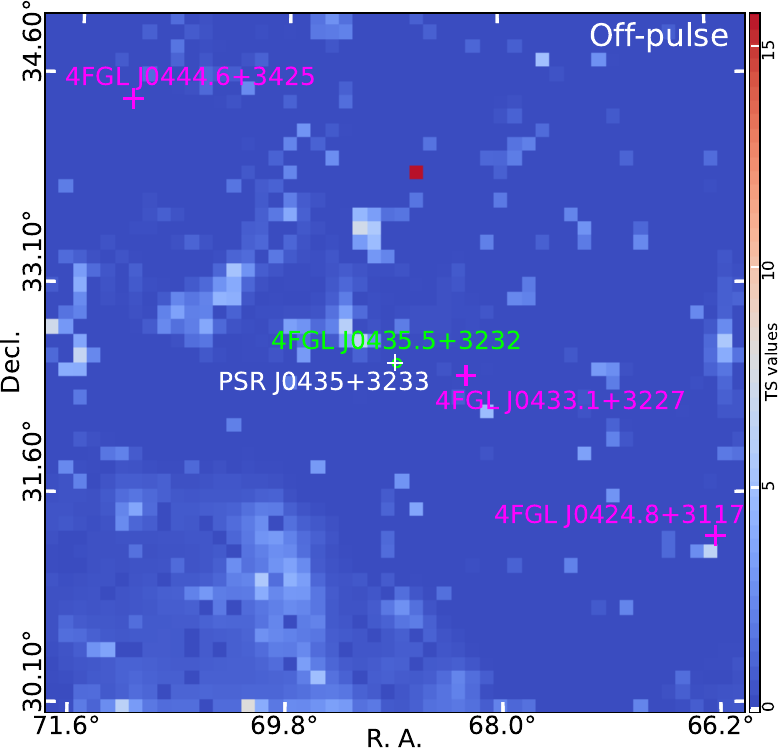}
\caption{TS maps of the $5^{\circ}\times5^{\circ}$ region centered on PSR~J0435+3233 in the
         0.1--500.0~GeV, constructed using events collected within the EVI. Left and right
         panels show the on-pulse and off-pulse TS maps, respectively. Both maps have a pixel
         size of $0.1^{\circ}\times0.1^{\circ}$. In each panel, the green circle indicates
         the $1\sigma$ positional uncertainty region of 4FGL~J0435.5$+$3232,
         and the white cross marks the radio timing position of PSR~J0435+3233 reported
         by \citet{wwy+26na}. The magenta crosses mark the positions of the other $\gamma$-ray
         sources listed in 4FGL-DR4.}
\label{fig:tsmap}
\end{figure*}

To illustrate the relative positions of 4FGL~J0435.5$+$3232 and PSR~J0435+3233, we also
constructed TS maps using the on-pulse and off-pulse data collected within the EVI,
as shown in Figure~\ref{fig:tsmap}. In each panel, the green circle represents
the $1\sigma$ positional uncertainty region of 4FGL~J0435.5$+$3232 \citep{4fgl-dr4}, while the white cross
marks the radio timing position of PSR~J0435+3233 reported by \citet{wwy+26na}.
The pulsar lies within the $1\sigma$ error circle of the 4FGL source.
And a $\gamma$-ray excess is visible near the pulsar position in the on-pulse TS map,
whereas no comparably significant excess is present in the off-pulse map. The spatial
and phase-dependent results therefore provide additional support for the association of
4FGL~J0435.5$+$3232 with PSR~J0435+3233 and for a predominantly pulsed origin of
the detected $\gamma$-ray emission.

\section{Summary and Discussion}
\label{sec:sd}

Using approximately 17.7~yr of \textit{Fermi}-LAT observations, we identify the cataloged source
4FGL~J0435.5$+$3232 as the $\gamma$-ray counterpart of PSR~J0435+3233. Restricting the timing
analysis to the EVI data, we detect $\gamma$-ray pulsations at a confidence level of $\sim6.8\sigma$.
The phase-resolved analysis shows that the detected emission is concentrated predominantly
within the rotational-phase interval $\phi\sim0.44$--$0.69$, which spans 0.25 in phase, while
no significant emission is detected in the remaining off-pulse interval.
The positional coincidence, the detection of pulsations at the pulsar's spin period,
and the phase-dependent emission collectively establish 4FGL~J0435.5$+$3232 as the $\gamma$-ray
counterpart of PSR~J0435+3233. Using the phase-averaged energy flux $G_{\gamma}$ measured over
the EVI, we calculate the $\gamma$-ray luminosity as $L_{\gamma}=4\pi f_{\Omega}d^{2}G_{\gamma}$,
where $f_{\Omega}$ is the beaming correction factor. Adopting $f_{\Omega}=1$ and a distance of
$d=1.2$~kpc, we obtain $L_{\gamma}=(6.26\pm2.48)\times10^{32}\ {\rm erg\,s^{-1}}$.
Despite its remarkably high spin-down luminosity of $\dot{E}=5.89\times10^{37}\ {\rm erg\,s^{-1}}$,
comparable to those of young, energetic pulsars \citep{wwy+26na}, PSR~J0435+3233 has an inferred
apparent $\gamma$-ray efficiency of only $\eta_{\gamma}=\frac{L_{\gamma}}{\dot{E}}\sim1.1\times10^{-5}$.

The combination of an exceptionally large spin-down power and an extremely low apparent
$\gamma$-ray efficiency makes PSR~J0435+3233 particularly unusual. The inferred efficiency
does not necessarily represent the intrinsic fraction of rotational energy converted into
$\gamma$-rays, because it depends on both the DM-derived distance and the unknown beaming
correction factor $f_{\Omega}$. The latter depends on the magnetic inclination,
viewing angle, and spatial distribution of the $\gamma$-ray emission within the magnetosphere.
The confinement of the detected emission to a limited rotational-phase interval
may be consistent with our line of sight sampling only a restricted portion of
the $\gamma$-ray beam, although the pulse width alone cannot uniquely constrain
the emission geometry. Moreover, the radio polarization profile cannot be adequately described
by the rotating-vector model and therefore provides no reliable constraints on
the magnetic inclination or viewing angle \citep{wwy+26na}. Joint modeling of absolutely
phase-aligned radio and $\gamma$-ray pulse profiles will be required to determine whether
the low apparent efficiency can be explained, at least in part, by the viewing and beaming geometry.

Alternatively, the low efficiency may be intrinsic, indicating that only a small fraction of
the available spin-down power is converted into GeV magnetospheric radiation.
The unusually strong surface dipole magnetic field of $B_{\rm s}\sim1.26\times10^{10}$~G for
an MSP, together with its millisecond rotation and potentially non-standard evolutionary history,
may give rise to a magnetospheric current configuration, accelerating electric field,
or pair-production environment different from those of typical recycled MSPs \citep{wwy+26na}.
In this case, a substantial fraction of the rotational energy could instead be carried away by
the pulsar wind or radiated outside the \textit{Fermi}-LAT energy range. PSR~J0435+3233 therefore
occupies a particularly interesting regime: it has the spin period and binary properties of an MSP,
but a spin-down power comparable to that of a young, energetic pulsar \citep{wwy+26na}.
Its 0.1--500.0~GeV properties may thus help clarify the relative roles of spin-down power,
magnetic and magnetospheric scales, viewing geometry, and evolutionary history in regulating
high-energy emission from pulsars.

To investigate whether the observed $\gamma$-ray emission contains a binary-related component
in addition to the pulsed magnetospheric emission, we also conducted an exploratory search for
variability at the binary orbital period using the Lomb--Scargle periodogram \citep{l76,s82}.
The current analysis does not provide sufficiently robust evidence to establish or characterize
orbital modulation, and we therefore refrain from using it to place strong constraints on the
binary-related $\gamma$-ray component. The detection of pulsations at the spin period securely
identifies a magnetospheric contribution, while the lack of significant off-pulse emission
indicates that this component dominates the detected signal. Nevertheless, the present GeV data
do not exclude a weaker contribution associated with an intrabinary shock,
interaction with the companion, or anisotropic emission in the binary environment.
The current $\gamma$-ray results also do not independently constrain the nature of the companion,
which may be a white dwarf, a low-mass main-sequence star, or an evolved star \citep{wwy+26na}.
Future $\gamma$-ray missions with improved sensitivity and angular resolution,
such as the Very Large Area Gamma-ray Space Telescope (VLAST; \citealt{fan+22}), may enable more
sensitive phase-resolved and orbital-variability studies, thereby providing tighter constraints on
the relative contributions of magnetospheric and binary-interaction-powered emission from this unusual MSP.

\begin{acknowledgments}
This work is supported in part by the National Natural Science Foundation of China under grant
Nos.~12233006 and 12163006. P.Z. acknowledges support from the Xingdian Talent Support Plan--Youth Project.
\end{acknowledgments}

\bibliographystyle{aasjournal}
\bibliography{aas}
\end{document}